\documentclass[pre,notitlepage,showpacs]{revtex4}

\newcommand{\Ap}{A_{\rm p}}
\newcommand{\eg}{e.g., }
\newcommand{\ie}{i.e., }
\newcommand{\kT}{k_{\rm B}T}
\newcommand{\pd}{\partial}
\newcommand{\pin}{p_{\rm in}}
\newcommand{\po}{p}
\newcommand{\vecq}{{\bf q}}

\begin{document}

\title{Model-free thermodynamics of fluid vesicles}

\author{Haim Diamant}

\affiliation{Raymond and Beverly Sackler School of Chemistry, Tel Aviv
University, Tel Aviv 69978, Israel}

\date{February 2, 2012}

\begin{abstract}
  Motivated by a long-standing debate concerning the nature and
  interrelations of surface-tension variables in fluid membranes, we
  reformulate the thermodynamics of a membrane vesicle as a generic
  two-dimensional finite system enclosing a three-dimensional volume.
  The formulation is shown to require two tension variables, conjugate
  to the intensive constraints of area per molecule and volume-to-area
  ratio. We obtain the relation between these two variables in various
  scenarios, as well as their correspondence to other definitions of
  tension variables for membranes. Several controversies related to
  membrane tension are thereby resolved on a model-free thermodynamic
  level. The thermodynamic formulation may be useful also for treating
  large-scale properties of vesicles that are insensitive to the
  membrane's detailed statistical mechanics and interactions.
\end{abstract}


\maketitle

\section{Introduction}
\label{sec_intro}

Membrane vesicles are flexible fluid envelopes, made of a bilayer of
amphiphilic molecules in aqueous solution \cite{Safran,Lipowski}.
Since Helfrich's work almost three decades ago \cite{Helfrich1973},
these prevalent structures have been one of the most extensively
studied systems in soft-matter physics, exhibiting rich
statistical-mechanical \cite{SeifertReview} and dynamic
\cite{Noguchi2007} behaviors. It may seem odd, therefore, to go back
now and reformulate their equilibrium thermodynamics.  Nevertheless,
this is the goal of the current work.  The reason is a long and
ongoing debate concerning various definitions of the surface
tension(s) of fluid membranes, their physical meanings, and
interrelations
\cite{Brochard1976,Peliti1985,David1991,Cai1994,Seifert1995,Fournier2001,Farago2003,Farago2004,Imparato2006,Stecki2008,Fournier2008,Fournier2008b,Barbetta2010,Schmid2011,Farago2011}.
Such issues as the exact number of independent tension variables and
their possible vanishing
\cite{Brochard1976,David1991,Farago2003,Schmid2011}, or even sign
reversal \cite{Fournier2008,Fournier2008b,Barbetta2010}, have remained
controversial. The debate has been conducted at the
statistical-mechanical level, emphasizing the role of thermal
fluctuations and employing various analytical and simulation
techniques. While certain issues are inherently
statistical-mechanical, it seems plausible that at least some of the
other fundamental issues mentioned above should be settled at the
lower, more general level of thermodynamics (\ie without specifying a
Hamiltonian).

Two key properties of fluid membranes make their thermodynamic
characterization special, and confusing. The first is that, while
being extensive in two dimensions, the membrane is embedded in three
dimensions. Consequently, it was suggested early on that the
thermodynamics of open membranes required two spatial constraints
rather than one \cite{Brochard1976}\,---\,the membrane area, $A$, and
its projection onto a reference planar frame, $\Ap$\,---\,entailing
two conjugate variables \cite{David1991}. The variable conjugate to
$A$, $\Pi$, is a surface pressure, analogous to ordinary thermodynamic
pressure. The one related to $\Ap$, the frame tension $\tau$, is the
force per unit length exerted at the edges of the frame, in the plane
of reference, to maintain a given $\Ap$.

The second property of bilayer membranes which sets them apart from
ordinary fluids is the fact that they have a preferred area per
molecule \cite{Israelachvili,Farago2003}. This boils down to the
hydrophobic effect, which opposes the exposure of the inner core of
the bilayer to the outer solvent \cite{Israelachvili}. As a result,
the membrane possesses in-plane elasticity\,---\,unlike an ordinary
fluid it can sustain not only a global compressive stress but also
a global dilative ones.

The discussion of membrane thermodynamics is further obscured when the
in-plane elasticity is suppressed by taking the limit of membrane
incompressibility. This commonly used assumption simplifies
calculations and allows one to theoretically fix the total area of the
membrane.  Yet, at the same time, it makes essentially independent
thermodynamic variables dependent\,---\,the area $A$ and number of
membrane molecules $N$ become linearly related, and so do the
conjugate surface pressure and chemical potential
\cite{David1991,Farago2003}.

To these complications, which are special to fluid membranes, one
should add the common pitfalls in thermodynamic analyses\,---\,the
necessity to precisely define which variables are kept fixed when
another variable is varied, and the fact that two variables, which
turn out to be equal under certain considerations, are not necessarily
identical.

To circumvent some of the foregoing complications we consider a closed
envelope\,---\,a vesicle\,---\,rather than an open membrane. We expect
(and will demonstrate below) that, once the thermodynamic behavior of
a vesicle is clarified, the consequences for an open membrane should
be readily inferred. In the case of a closed vesicle the
three-dimensional constraint is an ordinary volume constraint on the
enclosed system, rather than a projected-area constraint on the
surface. Thus, we explicitly consider the contents of the envelope as
a thermodynamic system, for which the properties of the membrane
become surface effects.  As a result, the starting point of the
formulation resembles that of other studies of finite-size
systems such as droplets or domains. It contains only well known,
unambiguous thermodynamic variables, such as volume and surface
area. The special properties of the membrane subsequently emerge
without further assumptions or an externally imposed reference
frame. In addition, membrane incompressibility is nowhere assumed.

The current work should be distinguished from other thermodynamic
formulations for membranes, which were presented in the past
\cite{Kozlov1989,Hoffmann2005}. The aim of those theories was to use a
{\em local} thermodynamic description, based on Gibbs' theory of
interfaces \cite{Kozlov1989} or molecular considerations
\cite{Hoffmann2005}, to construct a coarse-grained, local free energy
of a non-fluctuating mesoscopic element of the membrane. Subsequently,
the resulting free energy per unit area could be used, like the
Helfrich Hamiltonian \cite{Helfrich1973}, as a starting point for
statistical-mechanical or other calculations on a larger scale. By
contrast, we directly address the {\em global}, large-scale
thermodynamics of a membrane vesicle.

In Sec.\ \ref{sec_thermo} we formulate the thermodynamics of the
vesicle and present the resulting tension variables. In Sec.\ 
\ref{sec_scenario} we examine the consequences of the formulation in
various experimental and theoretical scenarios, and the resulting
relations between the two tension variables themselves, and between
them and other previously defined variables. While the formulation is
for a closed vesicle, we examine the analogous results for open
membranes as well. Section \ref{sec_discuss} summarizes the findings,
relates them with earlier results, and discusses their implications.

\section{Thermodynamic formulation}
\label{sec_thermo}

Consider a three-dimensional system of volume $V$, containing $Q$
particles \cite{ft_Qi} and being in thermal contact with a bath of
temperature $T$. In the thermodynamic limit the Helmholtz free energy
of the system, $F_3(T,V,Q)$, is extensive in $Q$ while keeping the
volume per particle fixed,
\begin{equation}
  F_3(T,V,Q) = Qf_3(T,v),\ \ \ v \equiv V/Q.
\label{F3}
\end{equation}
If we double the number of particles, and at the same time also double
the volume, the free energy $F_3$ will double. Conjugate to $T$ is the
entropy per particle, $s_3\equiv-(\pd f_3/\pd T)_{v}$. Conjugate to
$v$ is the internal pressure,
\begin{equation}
  \pin \equiv -\left(\frac{\pd f_3}{\pd v}\right)_T.
\label{p}
\end{equation}

The volume is enclosed by a two-dimensional envelope of surface area
$A$ and $N$ particles \cite{ft_smoothness}. As in studies of other
finite-size systems (\eg droplets, bubbles, domains), we restrict the
analysis to the leading correction introduced by the surface to the
thermodynamic limit $F=F_3$ of Eq.\ (\ref{F3}). This leading correction
arises from contributions to the free energy which are extensive in
the surface size $N$. Since the envelope is closed in three
dimensions, its Helmholtz free energy generally depends on both $A$
and $V$, $F_2=F_2(T,A,V,N)$. For example, the envelope's entropy
(which is extensive in $N$ and should be included, therefore, in
$F_2$) will decrease if $V$ is increased while keeping $A$ fixed. The
dependence of $F_2$ on $A$ and $V$ implies that the thermodynamic
description of the envelope requires {\em two} intensive variables
apart from temperature. We choose these variables to be the area per
particle, $a\equiv A/N$, and the dimensionless volume-to-area ratio,
$\alpha\equiv 6\sqrt{\pi}V/A^{3/2}$ (defined such that it has the
maximum value of unity for a perfectly spherical envelope). For $F_2$
to be extensive in $N$, one should fix {\em both} $a$ and $\alpha$,
\begin{equation}
  F_2(T,A,V,N)=Nf_2(T,a,\alpha),\ \ \ 
  a \equiv A/N,\ \ \   
  \alpha \equiv 6\sqrt{\pi}V/A^{3/2}.
\label{F2}
\end{equation}
This is a key point which is worth dwelling upon. Imagine, for
example, that we double both the number of particles making the
envelope and its area, but do not change the enclosed volume. The
surface free energy arising from in-plane short-range interactions and
in-plane entropy will double. Yet, evidently, the resulting object is
not merely a rescaled version of the original one. It is less
spherical and bound to have more out-of-plane fluctuation entropy per
molecule, leading to more than double the surface free energy. If we
double $N$ and $A$, and at the same time also increase $V$ by a factor
of $2^{3/2}$, the envelope will be properly rescaled. The in-plane as
well as the out-of-plane contributions per molecule will remain
unchanged, and the surface free energy $F_2$ will double. We stress
that Eq.\ (\ref{F2}) contains only contributions which are extensive
in the envelope's size. 

The appearance of the thermodynamic variable $\alpha$ reflects a
three-dimensional constraint imposed on the envelope. Here the
constraint is simply that the envelope be closed and contain a volume
$V$. One may impose other three-dimensional constraints, such as a
given projected area, $\Ap$, onto a reference surface.
Such a thermodynamic constraint is not independent of the volume
constraint; it may replace it but not be added to it. For example, if
the area $A$ of the envelope and the volume $V$ that it encloses are
known, then, evidently, the area of the sphere of volume $V$ (taken as
a reference surface), $\Ap=(4\pi)^{1/3}(3V)^{2/3}$, and its deviation
from $A$, are known as well. It is clear, however, that the foregoing
considerations of $A$ and $V$ are more transparent and independent of
any reference frame.

The flexibility of the thermodynamic formulation allows for adding
other constraints. For example, one may suggest that $F_2$ should
depend on additional characteristics of membrane shape apart from $A$
and $V$, such as the (average) curvature. Yet, one should examine
whether such additional constraints are beneficial or experimentally
relevant. It is hard to conceive a practical scenario where the shape
of a membrane is controlled to such an extent. We prefer, therefore,
to remain with the standard constraints of volume and area as
considered for any finite-size system. In statistical-mechanical terms
the membrane will sample all those additional shape characteristics
which are consistent with the global constraints of $A$ and $V$ (with
appropriate Boltzmann weights).

Returning to Eq.\ (\ref{F2}), we write the entropy per particle of the
envelope as $s_2\equiv-(\pd f_2/\pd T)_{a,\alpha}$.  The dependence of
$f_2$ on two intensive variables other than temperature, $a$ and
$\alpha$, implies that there are necessarily {\em two independent
conjugate variables}:
\begin{eqnarray}
  \Pi(T,a,\alpha) &\equiv& -\left(\frac{\pd f_2}{\pd a}\right)_{T,\alpha},
\label{Pi}\\
  \gamma(T,a,\alpha) &\equiv& \frac{3\alpha}{2a} 
   \left(\frac{\pd f_2}{\pd\alpha}\right)_{T,a}.
\label{gamma}
\end{eqnarray}
The prefactor in the definition of $\gamma$ has been introduced to
make the following presentation clearer. Both variables have
surface-tension dimensions. The first, $\Pi$, is the surface
pressure\,---\,the two-dimensional analogue of $\pin$\,---\,describing
the response of the envelope's molecules to in-plane strain (\ie
changes in the area per molecule without a change in the
volume-to-area ratio). At the optimum area per molecule $\Pi$
vanishes. The second variable, $\gamma$, referred to hereafter as the
Laplace tension, is related to the response of the envelope to
out-of-plane strain (\ie changes in the enclosed volume without a
change in the area per molecule).

We can further define three surface moduli,
\begin{eqnarray}
  K_{aa} &=& \left(\frac{\pd^2f_2}{\pd a^2}\right)_{T,\alpha} = 
   -\left(\frac{\pd\Pi}{\pd a}\right)_{T,\alpha}, \nonumber\\
  K_{\alpha\alpha} &=& \left(\frac{\pd^2f_2}{\pd \alpha^2}\right)_{T,a} = 
   \frac{2a}{3\alpha} \left[\left(\frac{\pd\gamma}{\pd\alpha}\right)_{T,a} 
   - \frac{\gamma}{\alpha} \right], \nonumber\\
  K_{a\alpha} &=& \frac{\pd^2f_2}{\pd\alpha\pd a} = 
   -\left(\frac{\pd\Pi}{\pd\alpha}\right)_{T,a} =
   \frac{2a}{3\alpha} \left[\left(\frac{\pd\gamma}{\pd a}\right)_{T,\alpha}
   + \frac{\gamma}{a} \right],
\label{moduli}
\end{eqnarray}
of which only the first is usually considered ($aK_{aa}$ is the
membrane's compression modulus). The last equality in Eq.\
(\ref{moduli}) is a Maxwell relation between the two variables, $\Pi$
and $\gamma$. In addition, convexity of $f_2$ demands that
$K_{aa}K_{\alpha\alpha}>K_{a\alpha}^2$.

Now consider the combined system, with total Helmholtz free energy
$F(T,V,Q,A,N)$. The fundamental differential relation for $F$
is
\begin{equation}
  dF = -SdT - \po dV + \mu_3dQ - \sigma dA + \mu_2 dN,
\label{dF}
\end{equation}
where $S$ is the total entropy, $\po$ the total pressure exerted by
the environment, $\sigma$ the total surface pressure in the envelope,
and $\mu_3$ and $\mu_2$ the chemical potentials, respectively, of the
particles enclosed inside the vesicle and those making the envelope.
In what follows we assume for simplicity that the particles in the
volume interact with those in the envelope via hard-core repulsion
only, \ie they are merely enclosed by the envelope. In this case the
total Helmholtz free energy is simply given by the sum of the volume
and surface contributions,
\begin{equation}
  F = F_3 + F_2 = Qf_3(T,v) + Nf_2(T,a,\alpha).
\label{F}
\end{equation}
Note, however, that there is an implicit coupling between the two
terms via their mutual dependence on the volume. Taking the
differential of Eq.\ (\ref{F}) and equating it with Eq.\ (\ref{dF}),
we identify:
\begin{eqnarray}
  S &=& Qs_3 + Ns_2, \\
  \po &=& \pin - \frac{2}{\alpha} \left(\frac{Na}{4\pi}\right)^{-1/2} \gamma,
\label{Laplace} \\
  \mu_3 &=& f_3 + \pin v, \\
  \sigma &=& \Pi + \gamma, 
\label{sigma} \\
  \mu_2 &=& f_2 + \Pi a.
\label{mu2}
\end{eqnarray}

Thus, the total surface pressure $\sigma$\,---\,the variable conjugate
to $A$ in the total free energy\,---\,is manifestly broken into two
distinct contributions, $\Pi$ and $\gamma$ [Eq.\ (\ref{sigma})]. We
underline the different roles played by each of these two tension-like
variables, defined in Eqs.\ (\ref{Pi}) and (\ref{gamma}).  The
similarity between $\Pi$ and ordinary thermodynamic pressure is
reflected in its relation to the chemical potential of the envelope's
molecules, Eq.\ (\ref{mu2}), in which $\gamma$ is absent.  The Gibbs
free energy of the envelope alone, whose sole extensive variable is
$N$, is $G_2=\mu_2N=F_2+\Pi A$ ({\em not} $F_2+\sigma A$). By contrast,
the mechanical nature of $\gamma$ is reflected in its contribution to
the pressure that the vesicle exerts on the external environment, Eq.\ 
(\ref{Laplace}), in which $\Pi$ does not appear. We may define for the
vesicle an effective thermodynamic radius of curvature, as
\begin{equation}
  R_{\rm c} \equiv \alpha\left(\frac{Na}{4\pi}\right)^{1/2}
  = \frac{3V}{A},
\label{Rc}
\end{equation}
which for a perfectly spherical envelope coincides with the sphere's
radius. With this definition Eq.\ (\ref{Laplace}) becomes Laplace's
law, $\pin-\po=2\gamma/R_{\rm c}$.

\section{Experimental and theoretical scenarios}
\label{sec_scenario}

As in any thermodynamic analysis, we can apply the formulation given
above to various scenarios, in which certain variables are controlled
while others are free to relax. These scenarios correspond to
different statistical ensembles. 
In the following subsections we describe several scenarios of
particular relevance to experimental conditions or to the debate over
membrane tension.

\subsection{Volume or pressure constraint}
\label{sec_volume}

Let us begin with a scenario in which we prescribe the volume enclosed
by the envelope, as well as the temperature and the number of
particles in the volume and in the envelope\,---\,\ie we fix
$(T,V,Q,N)$. In practice the volume may be considered fixed if we
assume  the inner solvent to be incompressible and neglect the
permeation of solvent into and out of the vesicle. (This is valid, for
example, over sufficiently short time.) Previous studies considered an
additional constraint of either a fixed area $A$ or a fixed total
surface pressure $\sigma$. We argue that this is unnecessary. Indeed,
it is impractical to dictate the area of the membrane. If we do not
impose an external surface pressure ($\sigma=0$; the case of finite
$\sigma$ is discussed in the next subsection), the area will relax
under the given constraints of $(T,V,Q,N)$ \cite{ft_rupture}.  The
Helmholtz free energy is then minimized with respect to $A$, leading to
\begin{equation}
  \Pi + \gamma = 0.
\label{balance}
\end{equation}
After the minimization, $F$ remains dependent only on $(T,V,Q,N)$. The
force balance expressed by Eq.\ (\ref{balance}) has the following
intuitive interpretation. The volume constraint acts as if an external
surface pressure were applied to the envelope, $\Pi_{\rm ex}=-\gamma$
\cite{ft_metric}. In response, the particles of the envelope develop
an internal surface pressure, $\Pi=\Pi_{\rm ex}$. The membrane's
response to out-of-plane strains must be restoring, $\gamma\geq 0$.
Hence, in the absence of an actual external surface pressure, the
in-plane balance is between two negative pressures\,---\,\ie the
volume constraint acts to increase the area, whereas the internal
pressure acts to decrease it.

Note that Eq.\ (\ref{balance}) is not an identity but an equation of
state, relating the envelope's intensive variables, $T$, $a$, and
$\alpha$; $\Pi$ and $\gamma$ are in general independent variables,
which become related in this scenario due to area relaxation. In
addition, using Eqs.\ (\ref{p}), (\ref{Pi}), (\ref{gamma}), and
(\ref{balance}), together with the known dependence of $\alpha$ on $V$
and $A$, one can calculate for a given model the unknown variables,
such as $\po$, $a$, $\Pi$, and $\gamma$, as functions of the
constraints $(T,V,Q,N)$ \cite{ft_no_A}.

Now consider a different case, where the pressure of the external
environment is controlled, while the volume can vary, \ie we fix
$(T,\po,Q,N)$. This scenario is quite similar to the preceding one; we
discuss it separately because it is the most relevant for actual
equilibrium vesicles, where the solvent has sufficient time to
permeate in and out of the vesicle. In this case we should minimize
the Gibbs free energy, $G\equiv F+\po V$, with respect to $V$ and $A$,
leading again to Eqs.\ (\ref{Laplace}) and (\ref{balance}). After the
minimization $G$ remains dependent only on $(T,\po,Q,N)$. We note
that, although we dictate $\po$, we do not directly fix $\pin$ because
of the surface effect represented by the Laplace tension $\gamma$ in
Eq.\ (\ref{Laplace}).

If there is neither a volume constraint nor a pressure one (\ie
$\po=\pin$), we have a tensionless membrane, $\Pi=\gamma=0$
\cite{David1991}.

\subsection{Contact with a reservoir of amphiphilic molecules}
\label{sec_mu2}

Let us examine a scenario in which the set of constraints is
$(T,\po,Q,\mu_2)$. We explicitly consider this case for two reasons.
First, it becomes the practically relevant scenario at sufficiently
long times, when the vesicle exchanges amphiphilic molecules with the
surrounding solution. Second, it highlights the crucial role played by
the three-dimensional embedding of the membrane. The latter issue is
recognized once we notice that the set $(T,\po,Q,\mu_2)$ does not
include any constraint that is extensive in membrane size. On the one
hand, such a thermodynamic formulation would normally be under-defined
and, hence, invalid. In the absence of any constraint on the
envelope's size, the vesicle can grow indefinitely by drawing more
amphiphilic molecules from the reservoir. On the other hand, it is
clear that this scenario can be realized in experiments and
simulations. This apparent contradiction is resolved below.

The free energy to minimize in this case is $\tilde{G}\equiv F+\po
V-\mu_2N$. Minimization of $\tilde{G}$ with respect to $V$, $A$, and
$N$ recovers Eqs.\ (\ref{Laplace}), (\ref{mu2}), and (\ref{balance}),
which we should be able to solve for $V$, $A$, and $N$, given the
constraints $(T,\po,Q,\mu_2)$. Let us explicitly rewrite the equations
in terms of these unknowns:
\begin{eqnarray*}
  0 &=& f_2\left(T,\frac{A}{N},\frac{V}{A^{3/2}}\right) + 
  \frac{A}{N} \Pi\left(T,\frac{A}{N},\frac{V}{A^{3/2}}\right) 
  - \mu_2,\\
  0 &=&  \Pi\left(T,\frac{A}{N},\frac{V}{A^{3/2}}\right)
  + \gamma\left(T,\frac{A}{N},\frac{V}{A^{3/2}}\right),\\
  0 &=& \pin\left(T,\frac{V}{Q}\right) 
  - \frac{2A}{3V} \gamma\left(T,\frac{A}{N},\frac{V}{A^{3/2}}\right)
  - \po.
\end{eqnarray*}
The first two equations, describing in-plane equilibrium, are
invariant to the scaling $(N\rightarrow\lambda N, A\rightarrow\lambda
A, V\rightarrow\lambda^{3/2}V)$ by an arbitrary scale factor
$\lambda$.  It is the third equation (Laplace's law), accounting for
the out-of-plane force balance, which violates this scaling and
prevents the set of equations from being under-defined.

\subsection{Surface pressure constraint}
\label{sec_sigma}

Just as it is impractical to impose a fixed area $A$ on a vesicle, it
is unclear how one could in practice apply to the vesicle a fixed
in-plane force per unit length, $\sigma$ \cite{ft_adhesion}.
Nonetheless, in statistical-mechanical studies of membranes such an
external surface-pressure term, $\sigma A$, is frequently added to the
Hamiltonian as a means to control the mean area. In the current
discussion there is no need to introduce such a term and, moreover, we
shall see that it leads to unnecessary complications; we consider it
merely to relate the emergent variables, $\Pi$ and $\gamma$, with the
commonly used $\sigma$. Within our thermodynamic formulation, such a
modification adds a term $\sigma A$ to the thermodynamic internal
energy and free energy \cite{ft_sigma}.  Consequently, minimization of
the free energy with respect to $A$ now yields Eq.\ (\ref{sigma}),
$\sigma=\Pi+\gamma$, as the equation of state instead of Eq.\ 
(\ref{balance}). One can again interpret this result as if the
particles of the envelope responded through an internal surface
pressure $\Pi$ to an external surface pressure, which consists of
$\sigma$ and the pressure $-\gamma$ due to the volume constraint,
$\Pi=\Pi_{\rm ex}=\sigma-\gamma$. Thus, just as prescribing $\po$ in
Sec.\ \ref{sec_volume} did not fix $\pin$ because of the existence of
$\gamma$, introducing the external surface pressure $\sigma$ does not
directly fix the internal one, $\Pi$, for a similar reason.

\subsection{Projected area and frame tension}
\label{sec_frame}

For open membranes a constraint on the membrane's projected area,
$\Ap$, onto a certain reference plane, is widely used. In the case of
a vesicle the choice of the reference surface itself is nontrivial and
may depend on the specific scenario \cite{Barbetta2010}. We emphasize
again that such a projection constraint is unnecessary in the current
analysis since, as noted in Sec.\ \ref{sec_thermo}, constraints on
$\Ap$ and $V$ are mutually dependent. We may introduce a projection
constraint {\it instead} of the volume constraint (on $V$ or $\po$) to
clarify the relations between the current formulation and previous
ones. Possible choices of a reference are the mean surface about which
the vesicle fluctuates \cite{Safran1983,Barbetta2010}, the spherical
shell that has the same area as the vesicle \cite{Seifert1995}, or the
spherical shell that holds the same volume as the vesicle
\cite{Milner1987,Seifert1995,Barbetta2010}. We choose the last option
as the reference surface, whose area is
\begin{equation}
  \Ap = \alpha^{2/3}A.
\label{Ap}
\end{equation}
We may also define the appropriate intensive variable, $a_{\rm
p}\equiv\Ap/N$.  Thus, fixing the projected area corresponds to fixing
neither $a$ nor $\alpha$ separately, but rather the product
$\alpha^{2/3}a$.

Related to the projected area is the frame tension, $\tau$. We have
not used the term `conjugate' intentionally, because the variable
conjugate to $\Ap$ would be the partial derivative of the membrane's
free energy with respect to $\Ap$ while keeping $T$, $N$, {\em and
$A$} fixed,
\begin{equation}
  \tau_1 \equiv \left(\frac{\pd F_2}{\pd\Ap}\right)_{T,A,N}
  = \left(\frac{\pd f_2}{\pd a_{\rm p}}\right)_{T,a}
  = \left(\frac{\pd f_2}{\pd \alpha}\right)_{T,a}
    \left(\frac{\pd \alpha}{\pd a_{\rm p}}\right)_{a}
  = \alpha^{-2/3}\gamma.
\label{tau1}
\end{equation}
However, the frame tension as measured or controlled in experiments
and simulations is related to the reversible work required to change
$\Ap$ while keeping only $T$ and $N$ fixed \cite{Farago2004}, as it is
impractical to fix $A$. This corresponds to differentiating $f_2$ with
respect to $a_{\rm p}$ while keeping only $T$ fixed; both $a$ and
$\alpha$ are allowed to vary while satisfying Eq.\ (\ref{Ap}) and the
equation of state $a=a(T,\alpha)$. (We have been assuming here the
absence of an external surface pressure; if a nonzero $\sigma$ is
considered, $\sigma$ should be kept constant as well.) This procedure
yields
\begin{eqnarray}
  \tau &\equiv& \left(\frac{\pd f_2}{\pd a_{\rm p}}\right)_{T}
  = \left(\frac{\pd f_2}{\pd\alpha}\right)_{T}
  \left(\frac{\pd\alpha}{\pd a_{\rm p}}\right)_{T}
  = \left[ \left(\frac{\pd f_2}{\pd\alpha}\right)_{T,a} +
  \left(\frac{\pd f_2}{\pd a}\right)_{T,\alpha} 
  \left(\frac{\pd a}{\pd\alpha}\right)_{T} \right]
  \left[ \frac{2}{3}\alpha^{-1/3}a + \alpha^{2/3}
  \left(\frac{\pd a}{\pd\alpha}\right)_{T} \right]^{-1}
 \nonumber\\
  &=& \alpha^{-2/3} \frac{\gamma-\epsilon\Pi}{1+\epsilon},\ \ \ \ \ \ \ 
  \epsilon \equiv \frac{3\alpha}{2a} \left(\frac{\pd a}{\pd\alpha}\right)_{T}.
\label{tau_general}
\end{eqnarray}

The general expression for $\tau$, Eq.\ (\ref{tau_general}), contains
a parameter $\epsilon$, which requires knowledge of the equation of
state $a(T,\alpha)$. This parameter vanishes for an incompressible
membrane having fixed $a$. It also cancels out in Eq.\ 
(\ref{tau_general}) if we use the equation of state (\ref{balance})
(\ie if we consider $\sigma=0$). In these two cases, therefore, we get
the simple relation,
\begin{equation}
  \mbox{$\sigma=0$ or incompressible membrane:}\ \ \ \ 
  \tau = \tau_1 = \alpha^{-2/3}\gamma = (A/\Ap)\gamma \geq \gamma.
\label{tau}
\end{equation}
Another simple limit is when the vesicle approaches a perfect
spherical shape (or the open membrane approaches a flat state), where
$\alpha\rightarrow 1$ and $\epsilon\rightarrow\infty$. Then we have
$\tau=-\Pi$, \ie the frame tension provides the entire pressure
required to balance the internal surface pressure. In the absence of
an external surface pressure, $\sigma=0$, this gives the expected
convergence for $\alpha=1$ of the frame and normal-response (Laplace)
tensions, $\tau=\gamma$.

If a nonzero $\sigma$ is considered, however, Eqs.\ (\ref{sigma}) and
(\ref{tau_general}) yield a different relation between the frame
tension and the other tension variables,
\begin{equation}
  \tau = \alpha^{-2/3} \left(\gamma - \frac{\epsilon}{1+\epsilon}\sigma \right),
\label{tau_sigma} 
\end{equation}
which requires model-dependent knowledge of $\epsilon$. The parameter
$\epsilon$ can also be represented in terms of the surface moduli
defined in Eq.\ (\ref{moduli}). Using the equation of state
(\ref{balance}) or (\ref{sigma}) to write $d(\Pi+\gamma)=0$ and
extracting from it the derivative $(\pd a/\pd\alpha)_{T}$, we obtain
\begin{equation}
  \epsilon = \frac{3}{2} \frac{(3/2)\alpha^2K_{\alpha\alpha} - a\alpha K_{a\alpha}
   + a\gamma} {a^2 K_{aa} - (3/2)a\alpha K_{a\alpha} + a\gamma}.
\label{epsilon}
\end{equation}
To assess the typical values of $\epsilon$ we make the following
estimates: we neglect the coupling modulus $K_{a\alpha}$; for a
vesicle approaching a spherical shape
$K_{\alpha\alpha}\sim\kT/(1-\alpha)^2$ \cite{HalevaPRL}, where $\kT$
is the thermal energy; $\gamma a$ does not exceed a few $\kT$ before
the membrane ruptures; $K_{aa}a^2$ is of order $10^2$ $\kT$. Thus, for
vesicles which are not extremely swollen, $\epsilon\ll 1$, and the
difference between Eq.\ (\ref{tau}) for $\sigma=0$ and
(\ref{tau_sigma}) for $\sigma\neq 0$ is small. For strongly swollen
vesicles ($1-\alpha<10^{-2}$), however, $\epsilon$ becomes large. In
this case we get $\tau\rightarrow\gamma-\sigma$.  Such a finite
difference between the frame tension and the Laplace tension as the
envelope and its reference surface coincide, is hard to physically
justify; it is clearly related to the artificial inclusion of an
external $\sigma\neq 0$.

\section{Discussion}
\label{sec_discuss}

The requirement that a membrane vesicle, as complicated as its
statistical mechanics may be, must comply with the thermodynamics of
finite-size systems, has yielded a surprising amount of information.
This information is in line with certain earlier findings and
disagrees with others.

The thermodynamics of a fluid vesicle requires two distinct tension
variables, defined here as the surface pressure $\Pi$ and the Laplace
tension $\gamma$ [Eqs.\ (\ref{Pi}) and (\ref{gamma})]. We did not
assume their existence but obtained them as a result of the standard
geometrical constraints on the area and volume of the vesicle. When
the area per molecule is allowed to relax\,---\,the relevant scenario
in actual systems and molecular dynamics simulations\,---\,the total
surface stress is zero, $\sigma=\Pi+\gamma=0$, and the two tension
variables become related. Unlike early suggestions
\cite{Brochard1976}, this does {\em not} imply that the membrane is
tensionless, but just reflects the balance between the surface stress
arising from a three-dimensional constraint and an equal internal
surface pressure. As already recognized in Ref.\ \cite{David1991}, a
state of vanishing tension, in the sense that the membrane does not
experience a normal restoring force due to tension, is where
$\gamma=0$. Then, upon area relaxation, we also have $\Pi=0$, \ie the
area per molecule relaxes to its optimum value.  This is achieved in a
vesicle, for instance, when its volume is allowed to relax in
conditions of zero pressure difference, $\pin=\po$ [Eq.\
(\ref{Laplace})]. A similar state is achieved in an open membrane
(unless a $\sigma\neq 0$ is introduced, or a non-optimum area per
molecule is imposed), when its projected area is allowed to relax
under conditions of zero frame tension, $\tau=0$ [Eq.\ (\ref{tau})].

Another tension coefficient that is frequently used in the literature
is the fluctuation tension, or `$q^2$-coefficient', $r$. It is
extracted from the membrane's normal fluctuation spectrum, $\langle
u(\vecq)u(-\vecq)\rangle = \kT/[rq^2 + O(q^4)]$, where $u(\vecq)$ is
the Fourier-transformed normal displacement of the membrane as a
function of wavevector $\vecq$. Evidently, this definition of tension
is purely statistical-mechanical and cannot be directly examined here.
It is clear, nonetheless, that $r$ is the tension related to the
membrane's response to normal strain. We thus suggest to identify it
as $r=\gamma$.  Indeed, micropipette-aspiration experiments have shown
that the measured $r$ is consistent with the imposed vesicle's Laplace
tension \cite{Evans1990}. For open membranes it was analytically
argued
\cite{David1991,Cai1994,Farago2003,Farago2004,Schmid2011,Farago2011},
and numerically demonstrated \cite{Farago2004,Neder2010,Farago2011},
that $r=\tau$, the frame tension, but other works disagree
\cite{Imparato2006,Stecki2006,Fournier2008,Sengupta2010}. Our analysis
confirms the former results. We have found that, when there is no
external surface pressure, $\sigma=0$, or when the membrane's
compressibility can be neglected, the Laplace tension $\gamma$ is
proportional to, but slightly smaller than, the frame tension $\tau$,
$\gamma=\alpha^{2/3}\tau=(\Ap/A)\tau$. This is in line with the
reports of $r=\tau$, since the difference
$(\tau-\gamma)u(\vecq)u(-\vecq)$ is only fourth-order in the normal
displacement.  It suffices to redefine the Fourier transform using the
membrane's material coordinates rather than its projected ones to get
a slightly modified spectrum with a $q^2$-coefficient
$r'=(\Ap/A)r=(\Ap/A)\tau=\gamma$. A recent numerical study suggests,
in fact, that the fluctuation tension is equal to $(\Ap/A)\tau$ rather
than $\tau$ \cite{Schmid2011}. As the vesicle becomes spherical (or an
open membrane gets planar) the frame tension coincides with the
Laplace tension, as expected.

The term $\sigma A$, added to the energy in many theories and
simulations, is the origin of a lot of confusion. Thermodynamically,
prescribing the variable $\sigma$ corresponds to applying an external
surface pressure \cite{ft_sigma}, similar to the $\po V$ term
introduced when a three-dimensional system is in contact with a
pressure bath. We are used to the fact that, for three-dimensional
systems, we must prescribe either the volume or the external pressure;
if we do not, there will be nothing to counterbalance the system's
internal pressure. One initially expects that for a membrane we would
similarly have to prescribe either $A$ or $\sigma$.  (The latter is
more widely used since studying fluctuations at fixed $A$ is
theoretically difficult \cite{Milner1987,Seifert1995}.)  However, in
an actual vesicle nothing fixes the area, and there is no in-plane
barrier that can apply a fixed external surface pressure
\cite{ft_adhesion}. Unlike the three-dimensional case, such an
external pressure is not necessary to balance the internal pressure
$\Pi$ of the two-dimensional fluid, because the volume constraint (\ie
the embedding of the membrane in three dimensions)\,---\,be it the
actual volume, a pressure difference, a projected area, or a frame
tension\,---\,provides such a balancing pressure, equal to $-\gamma$.
Even if we tune the volume constraint down to zero (\eg by imposing
$\pin=\po$ or $\tau=0$), the membrane can support a state of zero
internal surface pressure by attaining its optimum area per molecule,
without the need for an external pressure.  Thus, the area is
unconstrained, leading to $\Pi+\gamma=0$ \cite{ft_no_A}; equivalently,
we may say that the external surface pressure vanishes, $\sigma=0$. In
this restricted sense, the thermodynamics of area changes is similar
to the thermodynamics of photons, whose number is unconstrained or,
equivalently, whose chemical potential vanishes.

Hence, the ensembles of fixed $(T,V,Q,N)$ or $(T,\po,Q,N)$ with
$\sigma=0$ (Sec.\ \ref{sec_volume}) are the ones that we wish to
advocate as the most physically relevant and the least confusing. The
corresponding ensembles for open membranes are $(T,\Ap,N)$ or
$(T,\tau,N)$, again with $\sigma=0$. (At sufficiently long times the
constraint on $N$ should be replaced by one on $\mu_2$; see Sec.\
\ref{sec_mu2}.) The distinction between ensembles is not merely
technical or semantic, mainly because after removing the constraint on
$A$ or $\sigma$ the ensembles mentioned above contain one less
constraint than the ones usually studied. In addition, even ensembles
having the same number of constraints may not be equivalent for the
finite systems under consideration, which contain both volume and
surface free-energy contributions \cite{Schmid2011,Haleva2008}.

If one insists on including a nonzero $\sigma$ in the theory or
simulation, then area relaxation will lead to $\sigma=\Pi+\gamma$,
implying that $\sigma\neq\gamma=r$ in general.  The inequalities
$\sigma\neq r$ and $\sigma\neq\tau$ were observed in simulations
\cite{Fournier2008,Schmid2011,Farago2011}. We point out that the
numerically measured difference, $\sigma-\gamma=\sigma-r$, provides an
indirect measurement of the internal surface pressure $\Pi$ in such
nonzero-$\sigma$ computations. The inclusion of $\sigma\neq 0$ entails
two additional problems which, as we have been trying to show, are
superfluous. (a) Even when the membrane is tensionless, $\gamma=0$,
there is a residual surface stress, $\Pi=\sigma$, causing the area per
molecule not to relax to its optimum value. (b) The frame tension
$\tau$ is not proportional to $\gamma=r$ [Eq.\ (\ref{tau_sigma})].
Although the deviation should be usually small, when the vesicle is
strongly swollen (or an open membrane is nearly flat), we get
$\tau\rightarrow\gamma-\sigma$, which is physically questionable for
an envelope arbitrarily close to its projection.

We end by commenting on the possible signs of the various tension
variables \cite{Fournier2008,Fournier2008b,Barbetta2010}. The Laplace
tension, which characterizes the membrane's normal restoring force,
must be non-negative for the membrane to be stable, $\gamma\geq 0$.
This is reflected also in our identification $\gamma=r$, since a
negative $r$ will cause unstable large-wavelength fluctuations. If we
let the area relax without an external surface pressure,
$\sigma=0$\,---\,the conditions which, as we have argued, are the
valid ones for actual vesicles\,---\,then the condition $\gamma\geq 0$
entails a non-positive (compressive) internal pressure [Eq.\ 
(\ref{balance})], $\Pi\leq 0$, and a non-negative frame tension [Eq.\ 
(\ref{tau})], $\tau\geq 0$.  However, if a $\sigma\neq 0$ is imposed,
$\Pi$ can be also positive (dilative) [Eq.\ (\ref{sigma})], and $\tau$
can be also negative [Eq.\ (\ref{tau_sigma})].

The simple thermodynamic framework, laid out above, may be practically
useful in cases where the detailed statistical mechanics and
interactions of a membranal system are not crucial. For example, a
similar formulation has recently been used to unravel a universality
in the osmotic swelling of vesicles, arising from the mere competition
of volume and surface effects \cite{Peterlin2011}. There are issues
which obviously cannot be dealt with using thermodynamics alone.
Important examples are the fluctuation spectra, renormalization of
membrane parameters by fluctuations
\cite{Brochard1976,Peliti1985,Kleinert1986,David1991,Cai1994}, and the
equivalence (or lack of it) of different statistical ensembles
\cite{Schmid2011,Haleva2008}. Nonetheless, within its limitations, the
remarkable strength of classical thermodynamics lies in its
simplicity, broad applicability, and rigor \cite{Einstein}. We hope
that the formulation and conclusions presented here will be useful as
guidelines for future studies.


\begin{acknowledgments}
  I am thankful to Oded Farago for sharing his results prior to
  publication and for numerous discussions. Helpful discussions and
  correspondence with J.-B.\ Fournier, E.\ Haleva, M.\ Kozlov, J.\
  Landy, and F.\ Schmid are gratefully acknowledged.  Acknowledgment
  is made to the Donors of the American Chemical Society Petroleum
  Research Fund for partial support of this research (Grant No.\
  46748-AC6).
\end{acknowledgments}



\begin{thebibliography}{99}
\bibitem{ft_Qi} 
For simplicity we consider a single type of molecule, but the
formulation can be readily extended to a mixture of several species.
\bibitem{ft_smoothness}
No assumption concerning the smoothness of the envelope is necessary,
so long as one can calculate its area (\eg by considering the
particles as nodes and triangulating).
\bibitem{ft_no_A} 
The fact that the area constraint can be relaxed with $\sigma=0$
implies that we could, in principle, start from a surface free
energy $F_2$ which did not depend on $A$, assuming $A$ {\it a
priori} to be the relaxed area as a function of $(T,V,Q,N)$.  Such
a formulation, though equivalent to the one presented here (where we
introduce an area constraint and then relax it), would be much less
transparent.
\bibitem{ft_adhesion}
It might be possible to control the in-plane surface pressure of
vesicles by adhesion to a substrate at zero contact angle.
\bibitem{ft_sigma} 
In the statistical mechanics of membranes one distinguishes between
intrinsic tension\,---\,a coefficient that multiplies the area at
the ``microscopic'' level of the Hamiltonian\,---\,and renormalized
tension\,---\,the proportionality coefficient of $A$ in the free
energy. Evidently, this distinction lies beyond the scope of the
current thermodynamic discussion.
\end{thebibliography}
\end{document}